\documentclass[pra,twocolumn,superscriptaddress,longbibliography]{revtex4-2}
\usepackage[colorlinks=true, citecolor=blue, urlcolor=blue, linkcolor=blue ]{hyperref}
\usepackage{empheq,amssymb,orcidlink}

\begin{document}
\title{Protection of quantum steering ellipsoids in non-Markovian environments}
\author{Wen-Jie Zhang}
\affiliation{Key Laboratory of Quantum Theory and Applications of MoE, Lanzhou Center for Theoretical Physics, Gansu Provincial Research Center for Basic Disciplines of Quantum Physics, Key Laboratory of Theoretical Physics of Gansu Province, Lanzhou University, Lanzhou 730000, China}
\author{Jun-Hong An\orcidlink{0000-0002-3475-0729}}\email{anjhong@lzu.edu.cn}
\affiliation{Key Laboratory of Quantum Theory and Applications of MoE, Lanzhou Center for Theoretical Physics, Gansu Provincial Research Center for Basic Disciplines of Quantum Physics, Key Laboratory of Theoretical Physics of Gansu Province, Lanzhou University, Lanzhou 730000, China}
	
\begin{abstract}
Quantum steering ellipsoids (QSEs) provide a geometric representation, within the Bloch picture, of all possible states to which one qubit can be steered through measuring another correlated qubit. However, in realistic settings, quantum systems are inevitably coupled to their environment, resulting in decoherence and degradation of the QSE. Here, by investigating how local dissipative environments coupled to each qubit affect the quantum steering, we find that the geometry of each party’s QSE is closely tied to the non-Markovian effect and the formation of a bound state in the energy spectrum of the total qubit-environment system. The bound state provides the ability and the non-Markovian effect provides the dynamical way for preserving the QSE. We systematically examine the characteristics of QSEs under three distinct scenarios, i.e., two-sided bound states, one-sided bound states, and no bound state, revealing a diverse range of steering types. Our work establishes quantum reservoir engineering as a tunable strategy for protecting and controlling quantum steering in open systems, offering a practical pathway toward robust steering-based quantum technologies.
\end{abstract}
\maketitle
\section{Introduction}
Quantum steering is a quantum-correlation phenomenon in which one party, via local measurements, can remotely alter the state of another spatially separated party. First identified by Schr\"odinger \cite{Schrödinger_1935,Schrödinger_1936}, it has been formalized and extended through various frameworks \cite{GISIN1996151,HUGHSTON199314,PhysRevLett.98.140402,saunders_experimental_2010,PhysRevA.95.012320,PhysRevA.93.012108,PhysRevLett.130.221801,RevModPhys.92.015001,PhysRevLett.115.230402,PhysRevLett.114.060403,PhysRevA.40.913,PhysRevLett.106.130402,RevModPhys.89.015002,PhysRevLett.48.291,PhysRevLett.115.210401,PhysRevA.87.062103,PhysRevLett.122.240401,PhysRevA.101.042125,PhysRevX.5.041008,PhysRevLett.112.200402,PhysRevLett.113.050404,PhysRevLett.113.160402,PhysRevLett.70.1895,PhysRevLett.115.180502,PhysRevLett.122.070402,PhysRevA.109.032415}. For a two-qubit state, Alice’s steerability of Bob is fully captured by the quantum steering ellipsoid (QSE), which is formally defined as the whole set of Bloch vectors achievable for Bob's qubit by all possible positive-operator-valued measurements (POVMs) on Alice's qubit \cite{PhysRevLett.113.020402}. Just as the Bloch sphere offers an intuitive geometric picture for a single qubit \cite{Nielsen_Chuang_2010}, the QSE provides a faithful representation of a two-qubit state, where key quantum properties become directly visible in geometric terms, as exemplified by the nested tetrahedron condition for quantum separability \cite{PhysRevLett.113.020402, Milne_2014}.

The QSE captures the intricate geometry of two-qubit states, serving as a powerful visual tool that can simultaneously diagnose multiple forms of quantum correlation. Indeed, various features of the QSE have been employed to detect and quantify these different levels, including Bell nonlocality \cite{PhysRevA.90.024302}, Einstein-Podolsky-Rosen(EPR) steering \cite{PhysRevA.94.012114,Jevtic_2015,PhysRevA.95.012320,Song_2023,PhysRevA.97.022338,PhysRevA.95.042117}, entanglement \cite{PhysRevLett.113.020402,PhysRevA.94.012114,Milne_2014}, and quantum discord \cite{PhysRevLett.113.020402,PhysRevA.91.022301,PhysRevA.85.064104,Shi_2011}. Beyond these correlation hierarchies, the QSE has also proven powerful in characterizing other quantum phenomena, such as quantum phase transitions \cite{PhysRevA.104.012418,PhysRevB.107.085130}, maximal steered coherence \cite{hu_quantum_2016,PhysRevA.102.020402}, monogamous relations for entanglement and steering \cite{Milne_2014,PhysRevLett.122.070402,PhysRevA.94.042105}, and the experimental verification of QSEs themselves \cite{PhysRevA.109.032415,PhysRevLett.122.070402}. Such studies on QSE have traditionally focused on closed systems, neglecting environmental interactions. However, in realistic settings, quantum systems inevitably couple to their surroundings, leading to decoherence and the erosion of quantum features \cite{breuer_theory_2007}. It is therefore essential to develop methods that suppress decoherence in open quantum systems, thereby preserving the correlations captured by QSEs. Previous strategies for protecting QSE face a trade-off between effectiveness and resource overhead. Local operations can enhance discord but fail to maintain higher-order quantum correlations such as entanglement \cite{PhysRevA.91.022301}, whereas reservoir engineering with auxiliary qubits can successfully protect QSEs at the cost of extra resources \cite{PhysRevA.102.020402}. However, when these auxiliary qubits are absent, the system reverts to the Markovian regime and QSEs collapse rapidly. A scheme that can effectively protect QSE without auxiliary qubits would therefore be highly desirable. 

In this work we present an approach to QSE protection that operates without auxiliary qubits. We consider a bipartite quantum system in which each of Alice’s and Bob’s qubits interact with a local environment. We introduce a mechanism that suppresses environmental decoherence and thereby preserves the structure of the QSEs. Our analysis shows that this protection originates from the non-Markovian effect and the formation of bound states in the composite system of each qubit and its environment. We find that the preservation of the QSEs on each side is directly linked to the presence of such bound states. When bound states form on both sides, both QSEs maintain nontrivial structures which fail to satisfy the nested tetrahedron condition, thus confirming two-way quantum steering that is accompanied by entanglement. Our analysis further shows that two-sided bound states enable the recovery of two-way EPR steering, as quantified by standard EPR steering inequalities \cite{PhysRevA.93.012108}. Importantly, by adjusting the initial-state parameters, the same bound-state protection can be tailored to realize controlled one-way EPR steering. When a bound state exists on only one side, the QSE of that party is preserved and satisfies the nested tetrahedron condition, whereas the other party's QSE collapses to a point. This geometry corresponds to separable yet one-way steerable states. In the absence of any bound state, both QSEs decay and eventually collapse to a single point in the Bloch sphere. Matching the result under the Born-Markov approximation, this stands in sharp contrast to the protected QSEs sustained by bound-state formation in the non-Markovian dynamics.

The remainder of this paper is organized as follows. In Sec. \ref{1} we introduce the concept and significance of the QSE and discuss the classification of quantum steering. Section \ref{2} provides an analytical treatment of how the structure of the QSE in dissipative environments is linked to bound-state formation. In Sec. \ref{3} we present numerical results on the rich dynamical behaviors of the QSE under different bound-state scenarios, showing excellent agreement with analytical predictions. We summarize our main findings and conclusions in Sec. \ref{4}.

\section{QSE and steering Classification}\label{1}
Any two-qubit state shared between Alice and Bob may be expanded in the Pauli basis $\bm{\sigma} \equiv (\sigma_x, \sigma_y, \sigma_z)$ as $ \rho_{A B} = 1/4( \mathbb{I}_{A} \otimes \mathbb{I}_{B} + \mathbf{a} \cdot \boldsymbol{\sigma} \otimes \mathbb{I}_{B} + \mathbb{I}_{A} \otimes \mathbf{b} \cdot \boldsymbol{\sigma} + \sum\nolimits_{m, n=1}^{3} T_{mn} \sigma_{m} \otimes \sigma_{n} )$, where $\mathbb{I}_{A,B}$ are identity operators, $\mathbf{a}$ and $\mathbf{b}$ are the Bloch vectors of Alice's and Bob's reduced states, and $T$ is a $3\times3$ correlation matrix \cite{PhysRevA.54.1838}. In terms of components, $a_m = \mathrm{Tr}[\rho_{AB} \, \sigma_m \otimes \mathbb{I}_B]$, $b_n = \mathrm{Tr}[\rho_{AB} \, \mathbb{I}_A \otimes \sigma_n]$, and $T_{mn} = \mathrm{Tr}[\rho_{AB} \, \sigma_m \otimes \sigma_n]$. When Alice performs a local measurement on her qubit, each measurement outcome can be associated with a POVM element $\mathrm{E}\geq0$. Generally, $\mathrm{E}$ can be assigned to a Hermitian operator and thus we have $\mathrm{E}=e_{0}(\mathbb{I}_{A}+\mathbf{e}\cdot\boldsymbol{\sigma})$, where $0\leq e_0\leq1$ and $|\mathbf{e}|\leq1$. Bob is then steered to the unnormalized state $\operatorname{Tr}_{A}[\rho_{AB} \mathrm{E}\otimes\mathbb{I}_B]$ with probability $p=\operatorname{Tr}[\rho_{AB} \mathrm{E}\otimes\mathbb{I}_B]$. In particular, the normalized state admits the form $2^{-1}[\mathbb{I}_{B}+\left(\mathbf{b}+T^{\top}\mathbf{e}\right)\cdot\boldsymbol{\sigma}/(1+\mathbf{a}\cdot\mathbf{e})]$ for Bob’s qubit. Then, considering all possible local measurements by Alice, the set of Bob’s steered states corresponds to the following set of Bloch vectors \cite{PhysRevLett.113.020402,PhysRevLett.122.070402}
\begin{equation}
\mathcal{E}_B=\left\{\frac{\mathbf{b}+T^{\top}\mathbf{e}}{1+\mathbf{a}\cdot\mathbf{e}}:|\mathbf{e}|\leq 1\right\},\label{EA}
\end{equation}
It can be shown to describe an ellipsoid in the Bloch picture, referred to as Bob’s QSE $\mathcal{E}_B$ \cite{8603813,PhysRevLett.113.020402}. This ellipsoid is determined by its center $\mathbf{C}_{B}=\frac{\mathbf{b}-T^{\top} \mathbf{a}}{1-|\mathbf{a}|^{2}}$ and its ellipsoid matrix is given by
\begin{align} 
\mathcal{Q}_{B}=\frac{T^{\top}-\mathbf{b} \mathbf{a}^{\top}}{1-|\mathbf{a}|^{2}}\Big(\mathbb{I}+\frac{\mathbf{a} \mathbf{a}^{\top}}{1-|\mathbf{a}|^{2}}\Big)(T-\mathbf{a} \mathbf{b}^{\top}).
\end{align}
Similarly, Alice's steering ellipsoid $\mathcal{E}_{A}$ is generated when measurements are performed on Bob's qubit. Her QSE is centered at $\mathbf{C}_A = \frac{\mathbf{a} - T \mathbf{b}}{1 - |\mathbf{b}|^2}$ and her ellipsoid matrix is 
\begin{equation}
\mathcal{Q}_{A}=\frac{T-\mathbf{a b}^{\top}}{1-|\mathbf{b}|^{2}}\Big(\mathbb{I}+\frac{\mathbf{b b}^{\top}}{1-|\mathbf{b}|^{2}}\Big)(T^{\top}-\mathbf{b a}^{\top}).
\end{equation}
The eigenvalues of $\mathcal{Q}_{A}$ and $\mathcal{Q}_{B}$ give the squares of the semiaxis lengths of the respective ellipsoids, and their corresponding eigenvectors determine the orientations of those semiaxes \cite{PhysRevLett.113.020402}. Note that if the shared state is asymmetric under the exchange of Alice and Bob, the two steering ellipsoids are not identical. Together with Alice's and Bob's local Bloch vectors, the QSE provides a faithful geometric representation of the shared two-qubit state. As a powerful tool to reveal the nonclassical features inherent in the state, it provides a necessary and sufficient condition for entanglement through the nested tetrahedron condition that a two-qubit state is separable if and only if its QSE is enclosed by a tetrahedron inscribed in the Bloch sphere \cite{PhysRevLett.113.020402}. 

Quantum steering describes the phenomenon in which Alice can remotely prepare Bob's system in different states through appropriate local measurements \cite{Schrödinger_1936}. For two-qubit states, this steerability is captured by the geometry of the QSE, which contains all possible steerable states as described by Eq. \eqref{EA}. Quantum steering does not require entanglement and may be present in separable states \cite{PhysRevLett.113.020402}. On the other hand, there is another type of steering called EPR steering. It represents a stronger form of quantum correlation that sits between entanglement \cite{RevModPhys.81.865} and Bell nonlocality \cite{RevModPhys.86.419}. It reflects the ability of one observer to nonclassically influence the state of a remote system via local measurements, enabled by shared entanglement \cite{PhysRevLett.98.140402}. EPR steering is inherently directional, \cite{PhysRevLett.112.200402,handchen_observation_2012,PhysRevLett.116.160403,PhysRevLett.116.160404,PhysRevA.93.022121,PhysRevLett.112.180404,PhysRevA.88.051802,PhysRevA.81.022101,PhysRevResearch.4.013151,PhysRevA.109.022411,PhysRevLett.121.100401,PhysRevLett.131.110201,PhysRevLett.118.140404,PhysRevLett.130.200202}, which is distinct from  entanglement and Bell nonlocality and thus offers asymmetric advantages for quantum information tasks \cite{PhysRevLett.116.160403,PhysRevA.85.010301,gehring_implementation_2015}. Although the QSE for some special states determines the EPR steerability \cite{Jevtic_2015,quan_steering_2016,Nguyen_2016}, there is no direct connection between the EPR steering and the QSE for general two-qubit states. To certify EPR steering, we adopt the local uncertainty relation (LUR) as a criterion, using the three measurement settings $\{\sigma_{x,y,z}\}$ \cite{PhysRevA.93.012108,PhysRevLett.130.200202}. Suppose that Alice and Bob make measurements $A_i=\sigma_i$ and $B_i=\sigma_i$ on their respective qubits. Alice can steer Bob if the inequality $\sum_{i=x,y,z} \delta^{2} (\alpha_{i} {A}_{i}+{B}_{i}) \geq \min_{\rho_{B}} \sum_{i=x,y,z} \delta^{2} ({B}_{i})$ is violated, where $\delta^{2}(\cdot)$ denotes the variance of measurement outcomes and $\alpha_{i}=-(\langle {A}_{i}{B}_{i}\rangle-\langle{A}_{i}\rangle\langle {B}_{i}\rangle) / \delta^{2}(A_{i})$. For three-setting measurements, the right side satisfies $\min_{\rho_{A}}\sum_{i}\delta^{2}({A}_{i})=\min_{\rho_{B}}\sum_{i}\delta^{2}({B}_{i})=2$. Consequently, the steering witness $\Delta S_{AB}\equiv 2-\sum_{i}\delta^{2}(\alpha_{i} A_{i}+B_{i})$ becomes positive when Alice can EPR steer Bob. The corresponding criterion for Bob steering Alice, denoted by $\Delta S_{BA}$, follows an analogous form with the roles of Alice and Bob interchanged.

\section{QSE in noisy environments}\label{2}

We consider a scheme for protecting quantum steering of Alice and Bob, whose qubits are independently coupled to their own zero-temperature environments. This is a physical scenario where the two qubits are far apart. Thus, it is natural to consider that each qubit experiences local decoherence due to its interaction with its own environment. We follow the convention in open quantum systems that describes the environment causing the dissipative decoherence of qubit as a bosonic bath of harmonic oscillators. The Hamiltonian of the total system is $\hat{H} = \sum_{j=A,B} \hat{H}_j$, where
\begin{equation}
\hat{H}_j = \omega_{0} \hat{\sigma}_j^\dagger \hat{\sigma}_j + \sum_{k} [\omega_{k} \hat{b}_{jk}^{\dagger} \hat{b}_{jk} + ( g_{jk} \hat{\sigma}_j^\dagger \hat{b}_{jk} + \mathrm{H.c.} )].\label{e3}
\end{equation}
Here $\hat{\sigma}_j=|g_j\rangle \langle e_j|$ is the transition operator of the $j$th qubit from the excited state $|e_j\rangle$ to the ground state $|g_j\rangle$ with frequency $\omega_0$; $\hat{b}_{jk}^{\dagger}$ and $\hat{b}_{jk}$ are the creation and annihilation operators, respectively, of the $k$th mode with frequency $\omega_k$ in the environment coupled to the $j$th qubit. In the continuum limit, the spectral density of the environment is related to the coupling strength $g_{jk}$ as $J_j(\omega) \equiv \sum_{k}|g_{jk}|^{2} \delta(\omega-\omega_{jk})$. We consider that the spectral density $J_j(\omega)$ for the $j$th qubit explicitly takes the Ohmic family form $J_j(\omega) = \eta_j \omega^{s} \omega_{c}^{1-s} e^{-\omega / \omega_{c}}$, where $\eta_j$ is a dimensionless coupling constant, $\omega_c$ is the cutoff frequency, and $s$ is the Ohmicity parameter. The environment is classified as sub-Ohmic for $ 0<s<1$, Ohmic for $s=1$, or super-Ohmic for $s>1$. The Ohmic-family spectral density accurately captures the primary characteristics of the bosonic environment. First, the environment has infinite degrees of freedom, which are described by the Ohmic-family spectral density as a continuous function of $\omega$. Second, the environment's eigenenergy is bounded from below, which is also described by the Ohmic-family spectral density, with $J(\omega)=0$ for all frequencies below zero. Third, the interaction strengths of the environmental modes with the quantized light field are nonuniform, whose maximum is captured by the Ohmic-family spectral density as the mode with $\omega = s\omega_c$. Thus, the Ohmic-family spectral density is widely accepted for describing the interactions of a bosonic environment with open quantum systems, e.g. in the Caldeira-Leggett model of quantum Brownian motion and the spin-boson model \cite{RevModPhys.59.1,PhysRevD.45.2843}.

Considering the two environments are initially in the vacuum state, we can exactly trace out the environmental degrees of freedom from the unitary dynamics governed by Eq. \eqref{e3}. This yields a non-Markovian master equation for the reduced density matrix of the two-qubit system 
\begin{eqnarray}
&&\dot{\rho}_{AB}(t) = \sum_{j=A,B}\{-i\Omega_j(t)[\hat{\sigma}_{j}^{\dagger}\hat{\sigma}_{j},\rho_{AB}(t)]+\Gamma_j(t)\nonumber\\
&&~\times[2\hat{\sigma}_{j}\rho_{AB}(t)\hat{\sigma}_{j}^{\dagger}-\hat{\sigma}_{j}^{\dagger}\hat{\sigma}_{j}\rho_{AB}(t)-\rho_{AB}(t)\hat{\sigma}_{j}^{\dagger}\hat{\sigma}_{j}]\},\label{e5}
\end{eqnarray}
where $\Omega_j(t)=-\operatorname{Im}[\dot{c}_j(t) / c_j(t)]$ is the Lamb-shifted frequency and $\Gamma_j(t)=-\operatorname{Re}[\dot{c}_j(t) / c_j(t)]$ is the decay rate. The time-dependent coefficient $c_j(t)$ is determined by 
\begin{equation}
\dot{c}_j(t) + i\omega_{0}c_j(t) + \int_{0}^{t} f_j(t-\tau) c_j(\tau)  d\tau = 0, \label{e4}
\end{equation}
where $f_j(t-\tau)=\int_{0}^{\infty} J_j(\omega)e^{-i\omega(t-\tau)}d\omega$ is the correlation function of $j$th environment. The convolution kernel in Eq. \eqref{e4} captures the non-Markovian memory effect of the environment. 

Let us consider a two-qubit initial state shared between Alice and Bob as $\rho_{AB}(0)=p|\psi(\theta)\rangle\langle\psi(\theta)|+(1-p)\rho_{A}\otimes \mathbb{I}_{B}/2$, where $\theta \in (0, \pi/2)$, $p \in (0, 1)$, $|\psi(\theta)\rangle=\cos\theta|gg\rangle+\sin\theta|ee\rangle$, and $\rho_{A}=\operatorname{Tr}_{B}[|\psi(\theta)\rangle\langle\psi(\theta)|]$. In what follows, we omit the explicit time argument of $c_{A,B}(t)$ for brevity. According to Eq. \eqref{e5}, we obtain the reduced density matrix $\rho_{AB}(t) =\sum_{u,v;u',v'}\rho_{uu',vv'}|u,v\rangle\langle u',v'|$ in the energy eigenbasis $\{|gg\rangle, |ge\rangle, |eg\rangle, |ee\rangle\}$, where $\rho_{gg,gg} = 1 - |c_A|^2 \sin^2\theta+d$, $\rho_{ge,ge} = -d$, $\rho_{eg,eg} = -\frac{1}{2} |c_A|^2 [-2 + (1+p)|c_B|^2] \sin^2\theta$, $\rho_{ee,ee} = \frac{1}{2} |c_A c_B|^2 (1+p) \sin^2\theta$, and $\rho_{gg,ee}= \rho_{ee,gg}^*= c_A c_B \, p \cos\theta \sin\theta$, with $d=\frac{1}{2} |c_B|^2 [-1 + p \cos 2\theta + (1+p)|c_A|^2 \sin^2\theta]$, and all other elements are zero. 

Therefore, if Bob performs a measurement on his qubit, the center of Alice’s QSE is given by
\begin{equation}
\mathbf{C}_A=\left(0, 0, 1 + \frac{2  \sin^{2}\theta[(1 +p)(1+r|c_B|^{2})-r] }{r|c_A|^{-2} (2 + r|c_B|^{2})}\right),
\end{equation}
where $r=p\cos2\theta-1$ is negative for the parameter range considered. The semiaxis lengths $l_{A}^x,l_{A}^y$ and $l_{A}^z$ follow from the eigenvalues of $\mathcal{Q}_{A}$:
\begin{equation}
\begin{aligned}
& l_{A}^x=l_{A}^y=\frac{ p|c_A\sin 2\theta| }{ \sqrt{ -r(2 + r|c_B|^{2} ) } }, \\
& l_{A}^z= \frac{4p|c_A|^{2} \cos^{2}\theta \sin^{2}\theta }{ |r|(2 + r|c_B|^{2}) } .
\end{aligned}
\label{e7}
\end{equation}
In the same framework, the center and the lengths of the semiaxes of Bob's QSE are 
\begin{eqnarray}
   \mathbf{C}_B&=&\left(0, 0,1 - |c_B|^{2}(1+p) +\frac{p|c_B|^{2} \cos^{2}\theta}{1 - |c_A|^{2} \sin^{2}\theta} \right),\nonumber\\
   l_{B}^x& =&l_{B}^y=  \frac{ p|c_B \cos\theta|}{ \sqrt{1 - |c_A|^{2} \sin^{2}\theta } }, \label{e8}\\
 l_{B}^z&=& \frac{p|c_B|^{2} \cos^2\theta }{  1- |c_A|^{2} \sin^{2}\theta } .\nonumber
\end{eqnarray}

In the weak-coupling case where the timescale of the environment's correlation function $f_j(t-\tau)$ is much shorter than the characteristic timescale of the qubits, the Born-Markov approximation becomes applicable. Within this approximation, the solution of Eq. \eqref{e4} takes the form $c_j^{\mathrm{MA}}(t) \simeq e^{-[\kappa_j+i(\omega_{0}+\Delta_j)] t}$, with the decay rate $\kappa_j=\pi J_j(\omega_{0})$ and the Lamb shift $\Delta_j=\mathcal{P}\int_{0}^{\infty}\frac{J_j(\omega)}{\omega_{0}-\omega}d\omega$, where $\mathcal{P}$ denotes the Cauchy principal value. Substituting this into Eq. \eqref{e7} and \eqref{e8}, we readily obtain that the semiaxis lengths $l_A^{x,y,z}$ and $l_B^{x,y,z}$ of QSEs shrink to zero, indicating that all quantum correlations are lost over time. This result aligns with Markovian schemes, where the QSE collapses to a point in the absence of auxiliary qubits \cite{PhysRevA.102.020402}. 

Now let us go beyond the Born-Markov approximation. In the general non-Markovian case, Eq. \eqref{e4} must be solved numerically. Nevertheless, its long-time asymptotic behavior can be obtained analytically via the Laplace transform method. Specifically, Eq. \eqref{e4} is converted into $\tilde{c}_j(s)=(s+i\omega_{0}+\int_{0}^{\infty}\frac{J_j(\omega)}{s+i\omega}d\omega)^{-1}$ by Laplace transform. $c_j(t)$ is obtained by performing the inverse Laplace transform on $\tilde{c}_j(s)$, which requires finding the poles of  $\tilde{c}_j(s)$ via the equation 
\begin{equation}
Y_j(E) \equiv \omega_{0} - \int_{0}^{\infty} \frac{J_j(\omega)}{\omega - E}  d\omega = E, \quad (E = is).\label{S5}
\end{equation}
It is interesting to note that the root of Eq. \eqref{S5} precisely corresponds to the eigenenergy of each local qubit environment in the single-excitation subspace. To prove this, we expand its eigenstate as $|\Phi_j\rangle = ( x_j \hat{\sigma}_j^\dagger + \sum_k y_{jk} \hat{b}_{jk}^\dagger) |g_j, \{0_{jk}\} \rangle$. Substituting it into  $\hat{H}_j |\Phi_j\rangle = E_j |\Phi_j\rangle$, we obtain the energy eigenequation as $E_j - \omega_0 - \sum_k |g_{jk}|^2/(E_j-\omega_k ) = 0$, which matches Eq. \eqref{S5} in the continuous limit of $k$. Since $Y_j(E)$ decreases monotonically for $E<0$, Eq. \eqref{S5} admits a single isolated root $E_j^b$ in this region provided that $Y_j(0)<0$. In contrast, for $E>0$, $Y_j(E)$ loses analyticity, and Eq. \eqref{S5} possesses infinitely many roots that constitute a continuous energy band. We call the eigenstate of the isolated eigenenergy $E_j^b$ the bound state. Subsequently, applying the inverse Laplace transform and the residue theorem yields \cite{PhysRevA.103.L010601}
\begin{equation}
c_j(t)=Z_j e^{-i E_j^{b} t}+\int_{0}^{\infty} \frac{J_j(E) e^{-i E t} }{[E-\omega_{0}-\Delta_j(E)]^{2}+[\pi J_j(E)]^{2}}d E,
\end{equation}
where the first term with $Z_j \equiv \big[ 1 + \int_{0}^{\infty} \frac{J_j(\omega) \, d\omega}{(E_j^{b} - \omega)^{2}} \big]^{-1}$ originates from the bound state. The second term originates from the continuous energy band which decays to zero in the long-time limit due to the out-of-phase interference. Thus, we have
\begin{equation}
    \lim_{t\to\infty} c_j(t)=
    \begin{cases} 
    0; & \text{without bound state} \\ 
    Z_je^{-iE_j^{b}t}; & \text{with bound state}
    \end{cases}.\label{e9}
\end{equation}
The condition for bound-state formation in the Ohmic-family spectral density can be derived from $Y_j(0)<0$ as $\eta_j > \eta_{c} \equiv \omega_{0} / \omega_{c} \gamma(s)$, where $\gamma(s)$ denotes Euler’s gamma function. As an eigenstate of the total qubit-environment system in the single-excitation subspace whose eigenenergy resides outside the continuous energy band of the total system, the bound state arises when the qubit-environment coupling is so strong that the eigenstate becomes highly entangled. Its formation dynamically leads to the localization of the excitation in the qubit. In contrast, in the weak-coupling regime, all eigenstates are scattering states with energies residing in the continuous band, resulting in complete dissipation. Such a nonvanishing excitation amplitude of the qubits in the long-time limit due to the formation of the bound state is the key mechanism of the protection of QSE observed in our work.

Equations \eqref{e7} and \eqref{e8}, together with Eq. \eqref{e9}, clearly indicate the crucial role played by the bound state. In the absence of bound states, $c_j(t)$ decays to zero, causing the QSE to collapse to a single point within the Bloch sphere, just as in the Born-Markov case. In contrast, it is remarkable that the emergence of such bound states protects the semiaxis lengths from decay and sustains a finite QSE structure. Furthermore, the above equations reveal that the features of Alice's and Bob's QSEs $\mathcal{E}_A$ and $\mathcal{E}_B$ are strongly influenced by whether each of them individually forms a bound state. Specifically, distinct dynamical behaviors emerge depending on whether bound states are present on both sides, on one side only, or absent altogether. Consequently, the steady-state QSEs fall into distinct classes corresponding to these different bound-state configurations. Each of these cases will be examined in detail in what follows.

Similarly, using the LUR criterion together with Eq. \eqref{e9}, the EPR-steering witnesses $\Delta S_{AB}$ and $\Delta S_{BA}$ asymptotically tend to  
\begin{equation}
\begin{aligned}
\Delta S_{AB}(\infty) &= Z_A^2\{ h_A +p^2 Z_B^2 \cos[2(E_A^b+E_B^b)t] \sin^22\theta\}, \\
\Delta S_{BA}(\infty) &= Z_B^2\{ h_B +p^2 Z_A^2 \cos[2(E_A^b+E_B^b)t] \sin^22\theta\}.
\end{aligned}
\label{e10}
\end{equation} 
Here, the first terms $Z_j^2h_j$ are constants depending on $Z_j$, $p$, and $\theta$; the second terms describe oscillatory contributions. Explicit expressions for $h_A$ and $h_B$ are provided in the Appendix. Equation \eqref{e10} reveals that in the absence of bound states, i.e., $Z_{A,B}\to 0$, both terms vanish, thereby excluding any revival of EPR steering. This behavior matches the irreversible Born-Markov decay observed in noisy environments \cite{Rosales-Zarate:15}. When only Alice’s side lacks a bound state ($Z_A\to 0$), we obtain $\Delta S_{AB}(\infty)=0$ and $\Delta S_{BA}(\infty)<0$, due to $Z_B^2h_B$ reducing to $rZ_{B}^{2}(2-Z_{B}^{2}+Z_{B}^{2}p\cos 2\theta)$, which remains negative for all admissible parameters. Similarly, if only Bob’s side lacks a bound state ($Z_B\to 0$), we find $\Delta S_{BA}(\infty)=0$ and $\Delta S_{AB}(\infty)<0$, since $Z_A^2h_A$ becomes $4Z_{A}^{2}\sin^{2}\theta(-1+Z_{A}^{2}\sin^{2}\theta)$, which is also always negative. Hence, a one‑sided bound state alone cannot sustain the EPR steering. When both sides possess bound states, we can restore either two-way or one-way EPR steering by tuning the initial parameters $p$, $\theta$ and choosing suitable values of $Z_A$ and $Z_B$. This will be demonstrated in the following analysis.

It is noted that our scheme to protect the QSE requires two essential conditions: the existence of the bound state of the system and its environment and the non-Markovian effect. The bound state provides the ability and the non-Markovian effect provides the dynamical way of the QSE preservation. In sharp contrast to the transient enhancement of the QSE in Ref. \cite{PhysRevA.91.022301}, the QSE in our scheme is permanently present in the steady state. It has been reported that the QSE can be preserved by introducing auxiliary qubits to the reservoir \cite{PhysRevA.102.020402}. Without resorting to the auxiliary qubits, our scheme dramatically improves the experimental realizability.  

\section{numerical results}\label{3}
\begin{figure}
\includegraphics[width=\columnwidth]{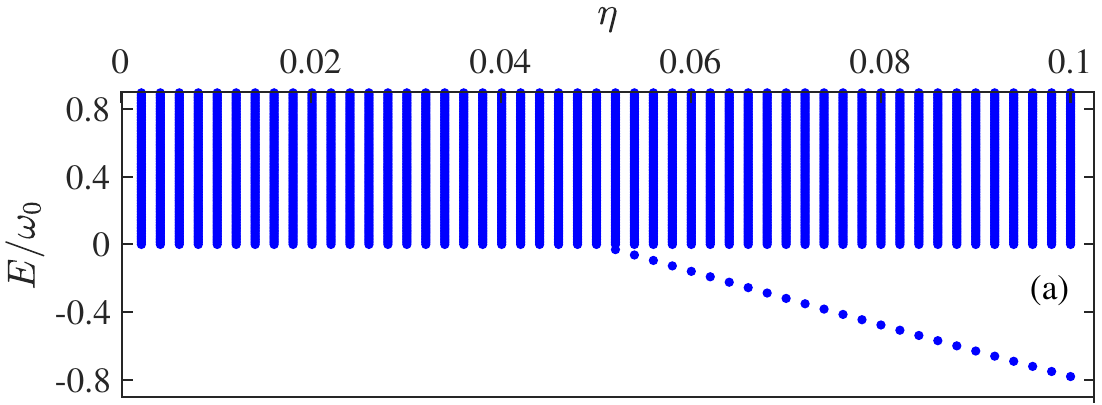}
\includegraphics[width=\columnwidth]{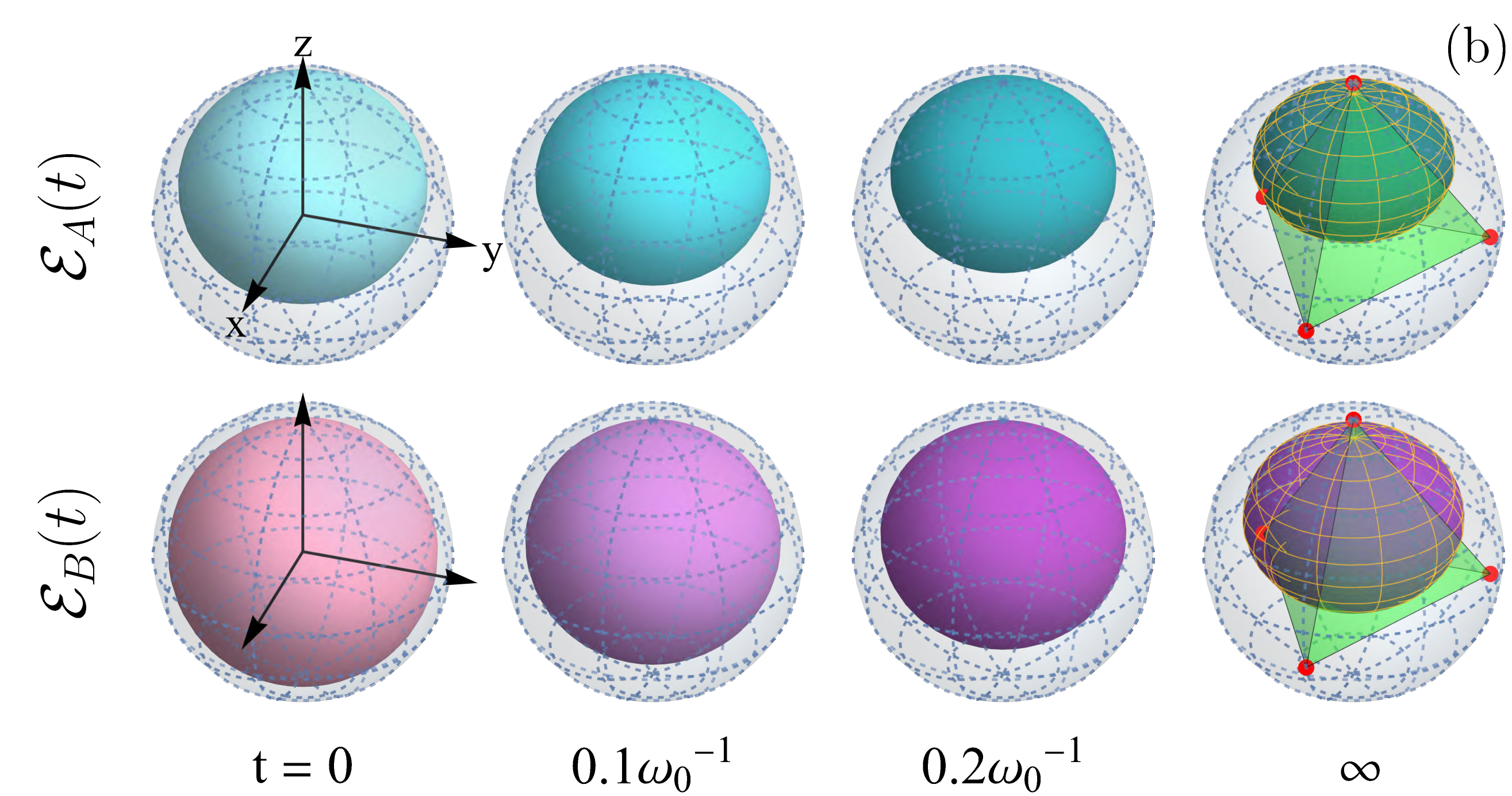}
\caption{(a) Energy spectrum of each qubit-environment subsystem. (b) Evolution of the QSEs $\mathcal{E}_A$ and $\mathcal{E}_B$ with $p=0.9$ and $\theta=\pi/8$. The solid surfaces depict the numerical evolution, with the analytical steady-state solution overlaid as a golden wireframe on the final ellipsoid. The green tetrahedron represents any tetrahedron inscribed in the Bloch sphere. It cannot fully enclose the steady-state steering ellipsoid. In all plots, we use $\omega_c=20\omega_0$ and $\eta_A=\eta_B=0.06$.} \label{FIG1}
\end{figure}

\begin{figure}
\includegraphics[width=\columnwidth]{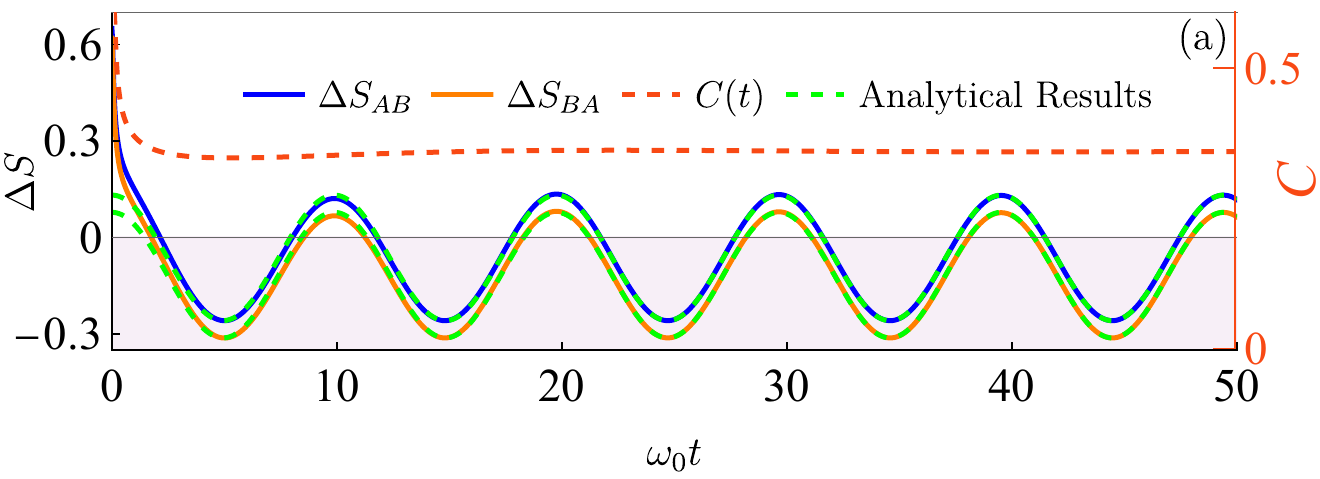}
\includegraphics[width=\columnwidth]{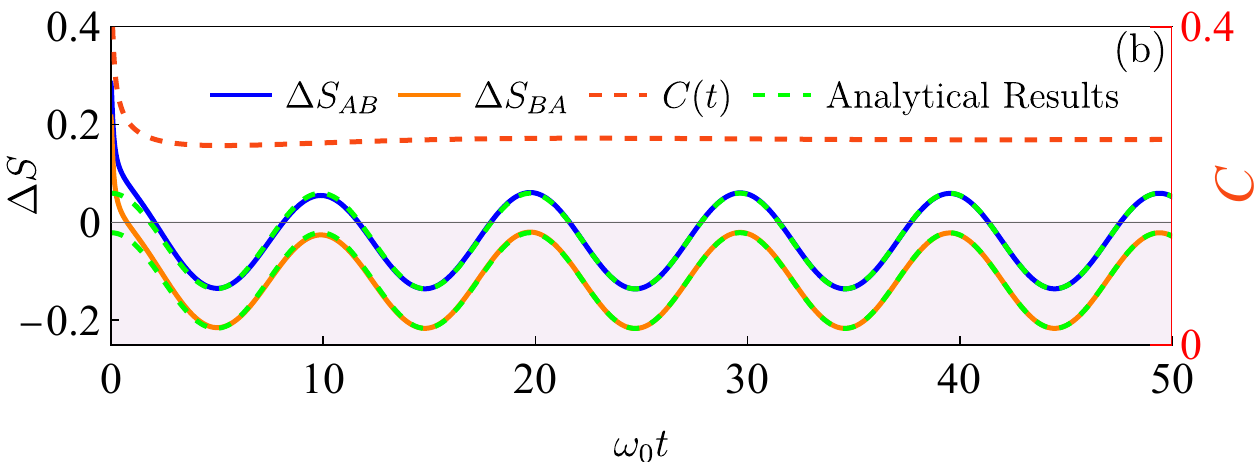}
\caption{(a) Evolution of the concurrence $C$ and the EPR steering witnesses $\Delta S_{AB}$ and $\Delta S_{BA}$ corresponding to the parameter set in Fig. \ref{FIG1}(a). Green dashed lines show the analytical results for $\Delta S_{AB}$ and $\Delta S_{BA}$ evaluated from Eqs. \eqref{e10}. Blue and orange solid curves are the corresponding numerical solutions. The red dashed line denotes the numerical evolution of the $C$. (b) Same as in (a) but with $\theta=\pi/12$.} \label{FIG2}
\end{figure}
To verify our analytical results, we choose an Ohmic spectral density, i.e., $s = 1$. For the initial state, we take the parameters $p=0.9$ and $\theta=\pi/8$, which support two-way EPR steering. Figure \ref{FIG1}(a) shows the energy spectrum of each qubit and its local environment. A bound state clearly emerges for $\eta>\eta_c=0.05$. Therefore, we set the coupling strengths on both sides to $\eta_A=\eta_B=0.06$, placing the two subsystems well within the bound-state regime. The solid red and purple surfaces in Fig. \ref{FIG1}(b) show the evolution of the QSEs $\mathcal{E}_A$ and $\mathcal{E}_B$, respectively, obtained from the exact numerical solution of Eq. \eqref{e4}. Notably, the QSEs converge to a steady robust structure in the long-time limit despite decoherence-induced contraction, contrasting sharply with the complete collapse observed without auxiliary qubits in Ref. \cite{PhysRevA.102.020402}. This steady structure is accurately captured by the analytical solution given in Eqs. \eqref{e7}, \eqref{e8}, and \eqref{e9}. The surface predicted by these equations is shown as a golden wireframe overlaid on the final solid QSE, aligning well with the converged numerical profile. Furthermore, we find that the steady-state QSE cannot be enclosed by any tetrahedron that fits inside the Bloch sphere, indicating that the state is entangled according to the nested tetrahedron condition \cite{PhysRevLett.113.020402}. These geometric features indicate that the steady state is entangled and supports two-way quantum steering. Meanwhile, entanglement is also quantified in Fig. \ref{FIG2}(a) with concurrence $C=\max \{0,\ \sqrt{\lambda_{1}} - \sqrt{\lambda_{2}} - \sqrt{\lambda_{3}} - \sqrt{\lambda_{4}}\}$, where $\lambda_l$ are the eigenvalues of $\rho_{AB}(\hat{\sigma}_{y}\otimes\hat{\sigma}_{y})\rho_{AB}^{*}(\hat{\sigma}_{y}\otimes\hat{\sigma}_{y})$ in decreasing order \cite{PhysRevLett.80.2245}. The persistence of $C$ at long times further confirms the nested tetrahedron condition as a reliable entanglement criterion.

\begin{figure}
\includegraphics[width=0.97\columnwidth]{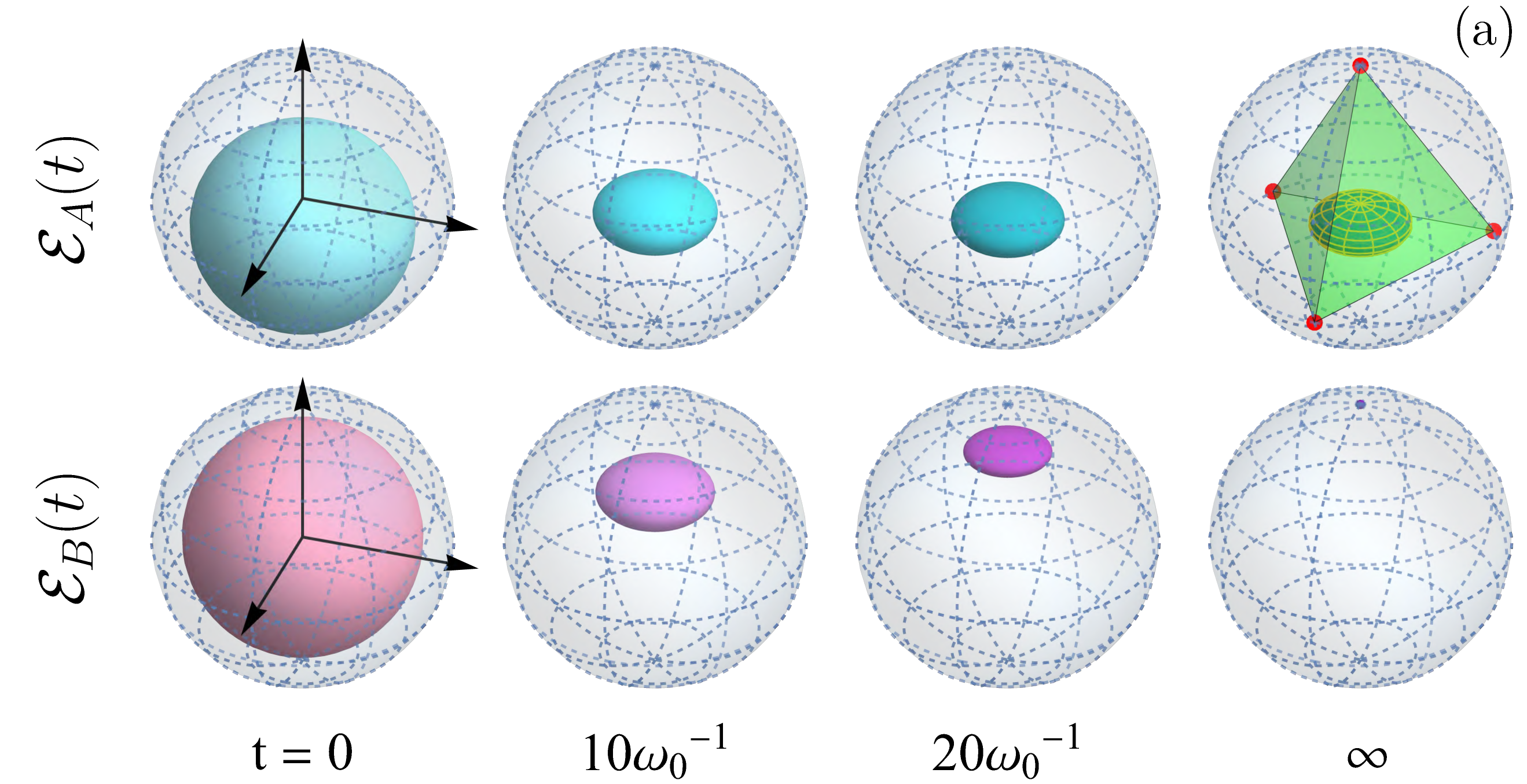}
\includegraphics[width=0.97\columnwidth]{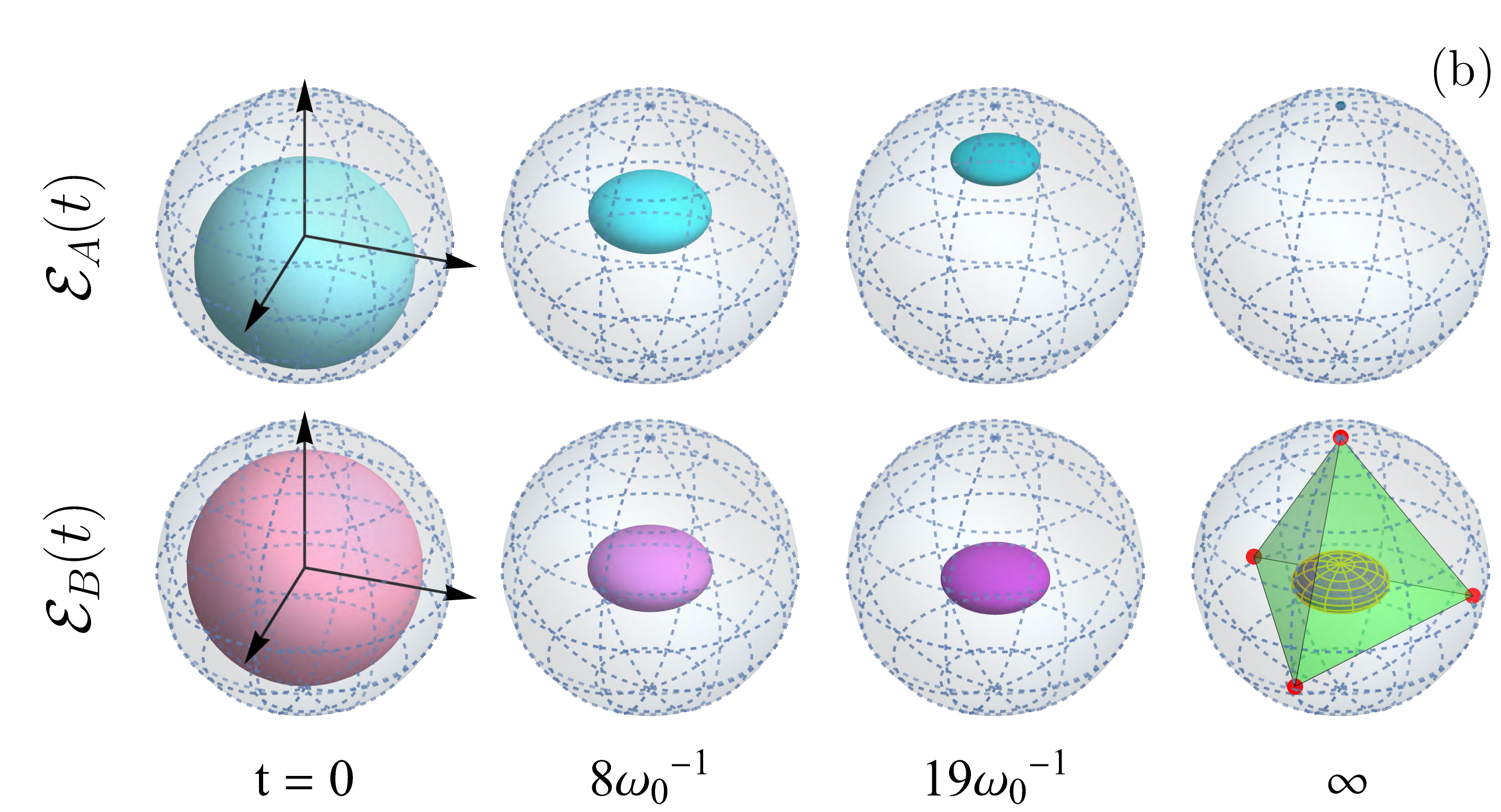}
\hspace*{9pt}\includegraphics[width=0.98\columnwidth]{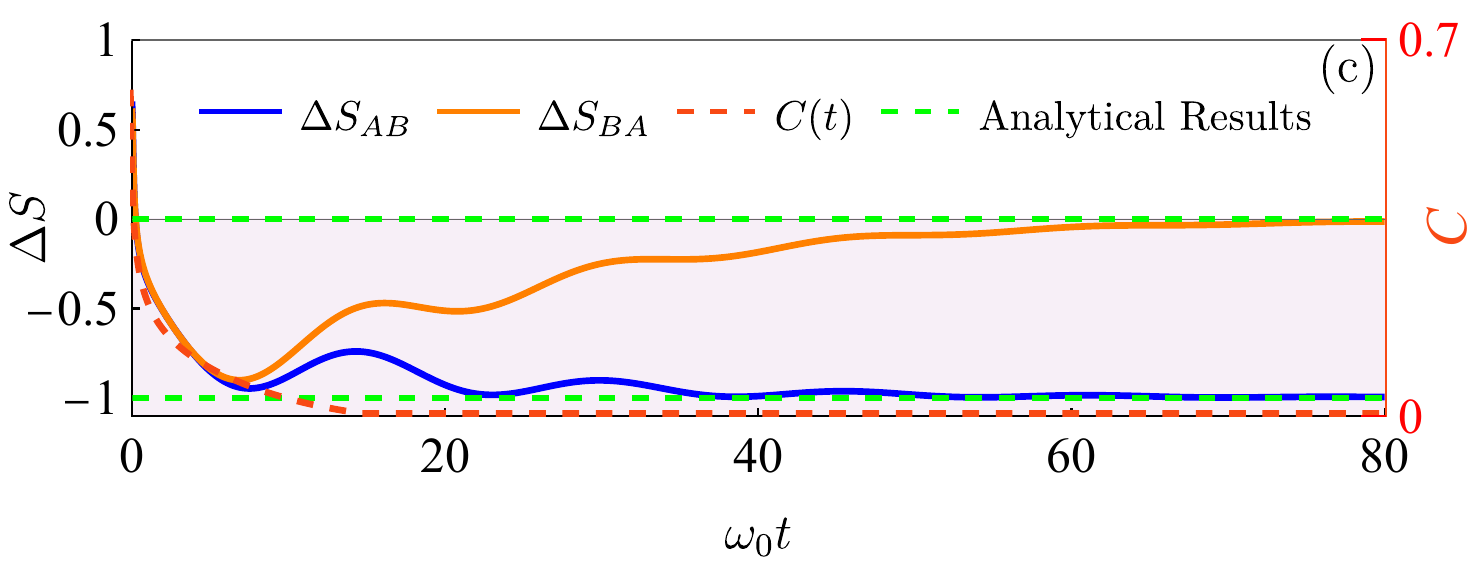}
\caption{(a) Evolution of the QSEs $ \mathcal{E}_A$ and $ \mathcal{E}_B$ with a bound state on Alice's side ($\eta_A=0.06$) and no bound state on Bob's side ($\eta_B=0.03$). (b) Reversed scenario with the bound state on Bob's side ($\eta_B=0.06$) and no bound state on Alice's side ($\eta_A=0.03$). The solid surfaces depict the numerical evolution, with the analytical steady-state solution overlaid as a golden wireframe on the final QSE. (c) Evolution of the $C$, $\Delta S_{AB}$ and $\Delta S_{BA}$ for the bound-state cases shown in (a). Green dashed lines show the analytical results for $\Delta S_{AB}$ and $\Delta S_{BA}$; blue and orange solid curves are the corresponding numerical solutions. The red dashed line denotes the numerical evolution of the $C$. In all plots, we use $\omega_c=20\omega_0$.} \label{FIG3}
\end{figure}

\begin{figure}
\includegraphics[width=0.97\columnwidth]{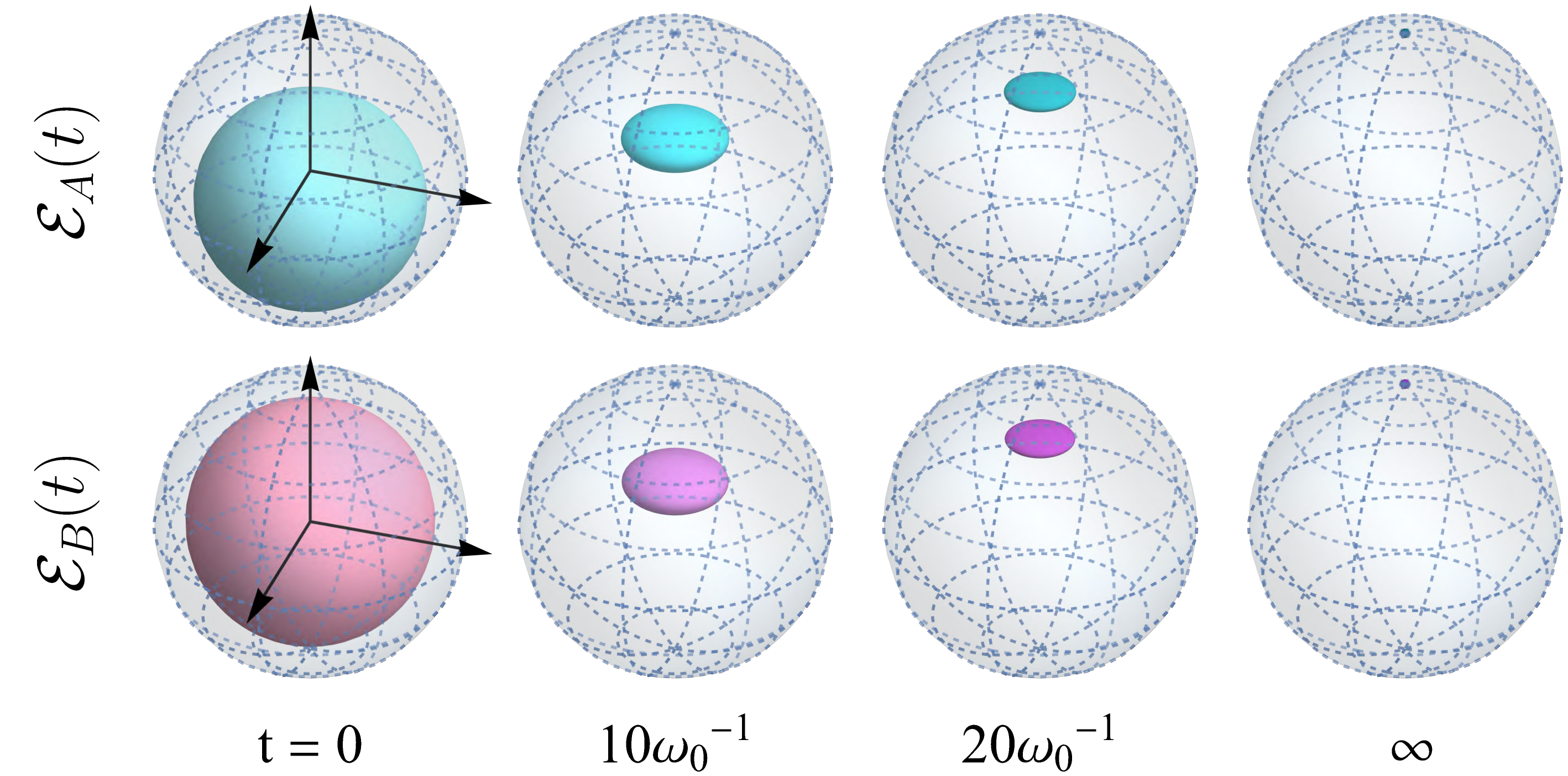}
\caption{Same as Fig. \ref{FIG3}(a) but with $\eta_A=\eta_B=0.03$, showing the evolution of the QSEs $ \mathcal{E}_A$ and $ \mathcal{E}_B$ with no bound state on either side.} \label{FIG4}
\end{figure}

To further test whether the steering represented by the QSE retains EPR-steering capability, we evaluate the numerical evolution of the EPR-steering witnesses $\Delta S_{AB}$ and $\Delta S_{BA}$ according to Eq. \eqref{e4}. As shown in Fig. \ref{FIG2}(a), their dynamics (solid curves) exhibit sustained oscillations with repeated sudden death and revival cycles, demonstrating the recovery of two-way EPR steering under bilateral bound‑state protection. The corresponding analytical results from Eq. \eqref{e10}, plotted as green dashed curves, closely match the numerical data. These sustained oscillations contrast with the Born-Markov approximate results in Ref. \cite{Rosales-Zarate:15}, where steering decays monotonically without revival. Notably, by adjusting the parameter $\theta$ in our chosen initial state, we can select between one-way and two-way EPR steering. This is demonstrated in Fig. \ref{FIG2}(b) for $\theta=\pi/12$, which shows the evolution of $\Delta S_{AB}$ and $\Delta S_{BA}$. In this case, $\Delta S_{BA}$ remains negative due to decoherence and shows no revival, while the intervals where $\Delta S_{AB}$ exceeds zero confirm the restoration of one-way EPR steering from Alice to Bob. Note that in this case the steady-state QSEs $\mathcal{E}_A$ and $\mathcal{E}_B$ (not explicitly plotted) still cannot be enclosed by any tetrahedron that fits inside the Bloch sphere, similar to Fig. \ref{FIG1}(b). This implies an entangled steady state that possesses the two-way quantum steering and one‑way EPR steering. A similar phenomenon has been found in Ref. \cite{PhysRevA.109.032415}. Moreover, as seen in Figs. \ref{FIG2}(a) and \ref{FIG2}(b), the EPR steering experiences sudden death during the evolution, whereas the entanglement $C$ remains finite in the system. This confirms that EPR steering represents a more stringent form of nonlocality than entanglement, requiring stronger correlations and being more susceptible to environmental disruption. The above results demonstrate that the bound states formed on both sides play a decisive role in protecting quantum steering from decoherence. Crucially, this protective mechanism supports both two-way and one-way EPR steering depending on the initial-state asymmetry, thereby offering a practical strategy to preserve and control steerable resources in open quantum systems. This capability is essential for applications such as quantum communication and asymmetric information protocols. We now discuss the effect of one-sided bound state on QSEs. Analytically, Eqs. \eqref{e7}, \eqref{e8} and \eqref{e9} reveal a clear relation between the semiaxes of the QSEs and the presence of a bound state in each subsystem. To verify this relation, we plot the numerical evolution of the QSEs $\mathcal{E}_A$ and $\mathcal{E}_B$ from Eq. \eqref{e4} for the initial state with $p=0.8$ and $\theta=\pi/3$. Figure \ref{FIG3}(a) shows the case where a bound state exists only on Alice's side, while Fig. \ref{FIG3}(b) corresponds to a bound state only on Bob's side. The numerically obtained steady-state QSEs agree well with the analytical predictions of Eqs. \eqref{e7}, \eqref{e8} and \eqref{e9}, whose surfaces are shown as golden wireframes, confirming the theoretical predictions. Our results show that, under decoherence, the QSE of the party possessing a bound state remains stable and maintains a well-defined structure in the long‑time limit. In contrast, the QSE of the party lacking a bound state shrinks continuously and eventually collapses to a single point on the Bloch sphere. Further geometric analysis shows that the surviving QSE lies entirely inside a tetrahedron inscribed in the Bloch sphere, which according to the nested tetrahedron condition indicates a separable state \cite{PhysRevLett.113.020402}. These geometric features ensure that the corresponding state, while being separable, still supports one-way quantum steering. The separability of the state is further evidenced by the complete suppression of concurrence $C$ shown in Fig. \ref{FIG3}(c). This well-defined one-way quantum steering stems directly from the ability of a one-sided bound state to provide asymmetric protection. Figure \ref{FIG3}(c) also presents the evolution of EPR steering witnesses $\Delta S_{AB}$ and $\Delta S_{BA}$ for the scenario shown in Fig. \ref{FIG3}(a). Their consistently non-positive behavior demonstrates that a one-sided bound state is insufficient to restore EPR steering, which stems from the loss of entanglement. It thereby offers a practical control strategy, where one can actively adjust the direction of quantum steering by engineering on which side the bound state appears. Finally, as a comparison, we present in Fig. \ref{FIG4} the numerical evolution of the QSEs $\mathcal{E}_A$ and $\mathcal{E}_B$ when neither side is protected by a bound state. The results show that in the absence of bound-state protection, both QSEs progressively degrade under decoherence and eventually collapse to a single point. This full collapse confirms the analytical predictions and is consistent with the Markovian results of \cite{PhysRevA.91.022301,PhysRevA.102.020402,Rosales-Zarate:15}. This comparison further verifies the indispensable role of bound states in sustaining non‑trivial steering capability and in enabling the design of one-way quantum steering.  

\section{Discussion and conclusion}\label{4}

It is emphasized that our bound-state based protection scheme for QSEs is independent of the specific choice of the spectral density. Although we have adopted the Ohmic-family spectral density, the underlying mechanism can be straightforwardly generalized to other forms without difficulty.  The quantum reservoir engineering techniques enable experimental control over the key parameters of the spectral density $J(\omega)$ \cite{PhysRevA.109.062402,andersen_strongly_2011,PhysRevB.95.161408,PhysRevA.84.031602}. The bound states and their associated dynamical effects have already been observed in state-of-the-art quantum optics experiments, including circuit quantum electrodynamics architectures \cite{liu_quantum_2017} and matter-wave systems \cite{krinner_spontaneous_2018}. This experimental progress provide strong support for our theoretical investigations. 

\begin{figure}
\includegraphics[width=1\columnwidth]{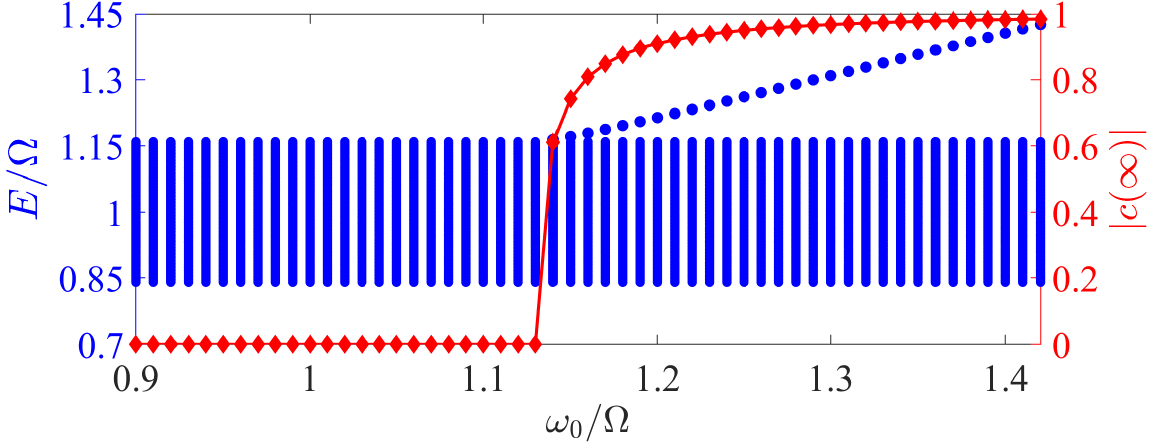}
\caption{Energy spectrum of a qubit coupled with an array of resonators for different $\omega_0$. The red diamonds represent the numerical results obtained from the long-time limit of Eq. \eqref{e4}. We use $g=0.05\Omega$ and $\xi=0.08\Omega$.} \label{R1}
\end{figure}

Our environment model can be experimentally realized using an array of coupled resonators, with implementations in microdisk cavity systems \cite{barclay_integration_2006}, optical resonator systems \cite{hafezi_imaging_2013}, optical waveguides \cite{PhysRevLett.110.076403,
rechtsman_photonic_2013}, and photonic crystal systems \cite{PhysRevB.86.195312,
hennessy_quantum_2007,
notomi_large-scale_2008}. The Hamiltonians of the environment and its interaction with the qubit are given by $\hat{H}_\mathrm{env} = \sum_{n=1}^{N-1} \Omega \hat{\tilde b}_{n}^\dagger \hat{\tilde b}_{n} + \sum_{n=1}^{N-1} \xi ( \hat{\tilde b}_{n}^\dagger \hat{\tilde b}_{n+1} + \mathrm{H.c.} )$ and $\hat{H}_{\mathrm{int}} = g ( \hat{\sigma}^\dagger \hat{\tilde b}_{1} + \mathrm{H.c.} )$, 
where $\hat{\tilde b}_n $ is the annihilation operator of the $n$th resonator with common frequency $\Omega$, $\xi$ is the coupling strength between nearest-neighbor resonators in the array, and $g$ is the coupling strength between the qubit and its own resonator array. The $\hat{H}_\mathrm{env}$ is rewritten in momentum space as $\hat{H}_{\mathrm{env}} = \sum_{k} \omega_k \hat{b}_{k}^\dagger \hat{b}_{k}$, where $\hat{b}_k$ is the Fourier transform of $\hat{\tilde{b}}_{n}$. The dispersion relation reads $\omega_k=\Omega+2\xi\cos{k}$, which yields a finite bandwidth of $4\xi$ centered at $\Omega$. The corresponding spectral density takes the form $J(\omega) = \frac{g^2}{2\pi\xi^2} \sqrt{4\xi^2 - (\omega - \Omega)^2}$. The band-gap structure in these platforms naturally supports the bound state. In Fig. \ref{R1} we plot the energy spectrum of the total system consisting of each qubit and its coupled resonator array, together with the steady-state value of $|c(\infty)|$ as a function of $\omega_0$. It is observed that a bound state emerges in the energy spectrum when $\omega_0$ exceeds a certain threshold. Correspondingly, $c(\infty)$ exhibits an abrupt transition from zero to a finite positive value at the critical point. According to Eqs. \eqref{e7} and \eqref{e8}, this nonzero steady-state amplitude directly leads to the preservation of the QSE structure.

In summary, we have proposed a scheme that protects QSEs in dissipative environments by exploiting bound-state formation and non-Markovian effect. This mechanism directly links the survival of a party’s QSE to the presence of a bound state on that side. When bound states exist on both sides, they protect the QSEs of both parties. As a result, the steady-state QSEs fail to satisfy the nested tetrahedron condition, thus signaling two-way quantum steering that is accompanied by entanglement \cite{PhysRevLett.113.020402}. This is further confirmed via the LUR criterion to be two-way EPR steering, exhibiting repeated periodic sudden death and revival in the non-Markovian dynamics. Notably, by tuning the initial-state asymmetry, the same protective mechanism can restore one-way EPR steering, a feature crucial for asymmetric quantum information tasks. If a bound state forms on only one side, the QSE of that party is preserved and satisfies the nested tetrahedron condition, while the other party’s QSE collapses to a point. This is a geometrical correspondence to separable states that exhibit one-way quantum steering. The resulting directional control offers a practical strategy: By engineering the side hosting the bound state, one can actively select the direction of quantum steering. Finally, when no bound state forms on either side, both QSEs collapse to a single point, as in the Born-Markov case of \cite{PhysRevA.91.022301,PhysRevA.102.020402,Rosales-Zarate:15}. These findings establish bound-state engineering as a versatile framework for protecting and controlling steering in open quantum systems, enabling the design of decoherence-resistant steering resources tailored to specific operational needs.

Our bound-state mechanism in preserving QSE is generalizable to multipartite systems. Explicitly, for a three-qubit state, one can compute pairwise reduced density matrices and construct the corresponding QSEs by tracing out the third qubit \cite{Milne_2014,PhysRevLett.122.070402}. When a bound state is present for the first qubit, the QSEs that describe the steerings of the second and third qubits to the first one would be preserved, respectively. In contrast, in the absence of bound states, the correlations between qubits would decay and eventually vanish.

\begin{acknowledgments}
The work was supported by the National Natural Science Foundation of China (Grants No. 12275109, No. 92576202, and No. 12247101), the Quantum Science and Technology-National Science and Technology Major Project (Grant No. 2023ZD0300904), the Gansu Science and Technology Leading Talent Program (Grant No. 26RCKA011 ), the Natural Science Foundation of Gansu Province (Grant No. 25JRRA799), and the Fundamental Research Funds for the Central Universities (Grant No. lzujbky-2025-jdzx07). 
\end{acknowledgments}

\begin{appendix}
\section*{APPENDIX: EXPLICIT EXPRESSIONS FOR $h_A$ AND $h_B$}\label{Appen}

The function $h_A$ is given by $h_A=h_{A1}-h_{A2}-h_{A3}$, where
\begin{align}
h_{A1} &= 4 \sin^2\theta(Z_B^2p^2\cos^2\theta-1), \\
h_{A2} &= \frac{4 Z_A^2 \sin^4\theta (2 + Z_B^2p^2 -Z_B^2 ) }{r(2 +r Z_B^2 )},\\
h_{A3} &= \frac{8 pZ_A^2 \sin^4\theta\cos 2\theta(  Z_B^2  +  p Z_B^2 - 1)  }{r(2 +r Z_B^2 )}.
\end{align}
The function $h_B$ is given by $h_B=h_{B1}+h_{B2}+h_{B3}$, where
\begin{align}
h_{B1} & = \frac{ Z_B^2 (1+ p^2 \cos^2 2\theta) - 2}{1 - Z_A^2 \sin^2\theta}, \\
h_{B2} & = Z_{A}^{2}p^{2}\sin^{2}2\theta+\frac{Z_{A}^{2}\left(2-Z_{B}^{2}+Z_{B}^{2}p^{2}\right)}{(1-Z_{A}^{2}\sin^{2}\theta)\sin^{-2}\theta}, \\
h_{B3} & =\frac{ 2p  \{1 - Z_B^2 + Z_A^2 [ Z_B^2 (1 + p)-1] \sin^2\theta\}}{(1 - Z_A^2 \sin^2\theta)\cos^{-1} 2\theta}.
\end{align}
\end{appendix}

\bibliography{ref}
\end{document}